\documentclass{emulateapj}
\usepackage{apjfonts}

\begin{document}
 
\title{The Thermal Evolution following a Superburst on an Accreting
Neutron Star}
 
\author{Andrew Cumming\altaffilmark{1} and Jared Macbeth}
\affil{Department of Astronomy and Astrophysics, University of
California, Santa Cruz, CA 95064;
cumming@ucolick.org}\altaffiltext{1}{Hubble Fellow}
 
\begin{abstract}
Superbursts are very energetic Type I X-ray bursts discovered in
recent years by long term monitoring of X-ray bursters, and believed
to be due to unstable ignition of carbon in the deep ocean of the
neutron star. In this Letter, we follow the thermal evolution of the
surface layers as they cool following the burst. The resulting light
curves agree very well with observations for layer masses in the range
$10^{25}$--$10^{26}\ {\rm g}$ expected from ignition calculations, and
for an energy release $\gtrsim 10^{17}$ erg per gram during the
flash. We show that at late times the cooling flux from the layer
decays as a power law $F\propto t^{-4/3}$, giving timescales for
quenching of normal Type I bursting of weeks, in good agreement with
observational limits. We show that simultaneous modelling of
superburst lightcurves and quenching times promises to constrain both
the thickness of the fuel layer and the energy deposited.
\end{abstract}


\section{Introduction}

Type I X-ray bursts from accreting neutron stars in low mass X-ray
binaries involve unstable thermonuclear burning of accreted hydrogen
(H) and helium (He) (Lewin, van Paradijs, \& Taam 1995). In the last
few years, long term monitoring of X-ray bursters by BeppoSAX and the
Rossi X-Ray Timing Explorer (RXTE) has revealed a new class of very
energetic Type I X-ray bursts, now known as "superbursts" (see
Strohmayer \& Bildsten 2003; Kuulkers 2003 for reviews).  The
$10^{42}\ {\rm erg}$ energies and several hour durations of
superbursts are $100$--$1000$ times greater than usual Type I
bursts. In addition, they are rare: so far 8 have been seen from
7 sources, with recurrence times not well-constrained, but
estimated as $\sim 1$ year (Kuulkers 2002; in 't Zand et al.~2003;
Wijnands 2001), instead of hours to days for usual Type I bursts.

The current picture is that superbursts are due to unstable ignition
of carbon at densities $\rho\sim 10^8$--$10^9\ {\rm
  g\ cm^{-3}}$. Hydrogen and helium burn at $\rho\sim
10^5$--$10^6\ {\rm g\ cm^{-3}}$ via the rp-process (Wallace \& Woosley
1981), producing chiefly heavy elements beyond the iron group
(including nuclei as massive as $A = 104$; Schatz et al. 2001), but
with some residual carbon (mass fraction $X_C\sim 0.01$--$0.1$)
(Schatz et al.~2003). Cumming \& Bildsten (2001) (hereafter CB01)
showed that this small amount of carbon can ignite unstably once the
mass of the ash layer reaches $\sim 10^{25}\ {\rm g}$ (see also
Strohmayer \& Brown 2002). This fits well with observed superburst
energies for $X_C\approx 0.1$ and an energy release from the nuclear
burning of 1 MeV per nucleon. The low thermal conductivity of the
rp-process ashes gives a large temperature gradient and ignition at
the required mass (CB01). The heavy nuclei may also photodisintegrate
to iron group during the flash, enhancing the nuclear energy release
(Schatz, Bildsten, \& Cumming 2003). Therefore superbursts offer an
opportunity to study the rp-process ashes.

Previous authors used one-zone models to estimate the time-dependence
of the flash (CB01; Strohmayer \& Brown 2002). In this paper, we
present the first multi-zone models of the cooling phase of
superbursts. Unlike normal Type I bursts, the time to burn the fuel is
much less than the convective turnover time.  We therefore assume that
the fuel burns locally and instantaneously in place, without
significant vertical mixing. We do not calculate ignition conditions,
but rather treat the amount of energy deposited and the thickness of
the fuel layer as free parameters\footnote{This is a similar approach
  to Eichler \& Cheng (1989) who studied the thermal response of a
  neutron star to energy deposition at different depths.  However, the
  transient events they consider are less energetic than
  superbursts.}. In \S 2, we describe our calculations of the
subsequent thermal evolution of the layer, and present a simple
analytic model which helps to understand the numerical results. At
late times, the flux evolves as a power law in time rather than the
exponential decay found by CB01 for a one-zone model. In \S 3, we use
the long term flux evolution of the layer to predict the timescale of
quenching of Type I bursts after the superburst, and compare to
observations.

\section{Time Evolution of the Superburst}

After the fuel burns, the cooling of the layer is described by the
entropy equation
\begin{equation}
c_P {\partial T\over\partial t} = -\epsilon_\nu -{1\over\rho}{\partial
F\over \partial r}
\end{equation}
where the heat flux is $F = -K(\partial T/\partial r)$, and
$\epsilon_\nu$ is the neutrino energy loss rate. The layer remains in
hydrostatic balance, in which case a useful independent coordinate is
the column depth $y$ into the star (units: ${\rm g\ cm^{-2}}$), where
$dy = - \rho dr$, giving a pressure $P =gy$. The surface gravity is $g
= (GM/R^2)(1 + z)$, where $1 + z = (1 - 2GM/Rc^2)^{-1/2}$ is the
gravitational redshift factor. In this paper, we assume $M = 1.4
M_\odot$ and $R = 10 {\rm km}$, giving $z = 0.31$ and $g_{14} =
g/10^{14}\ {\rm cm\ s^{-2}} = 2.45$.

\begin{figure}
\plotone{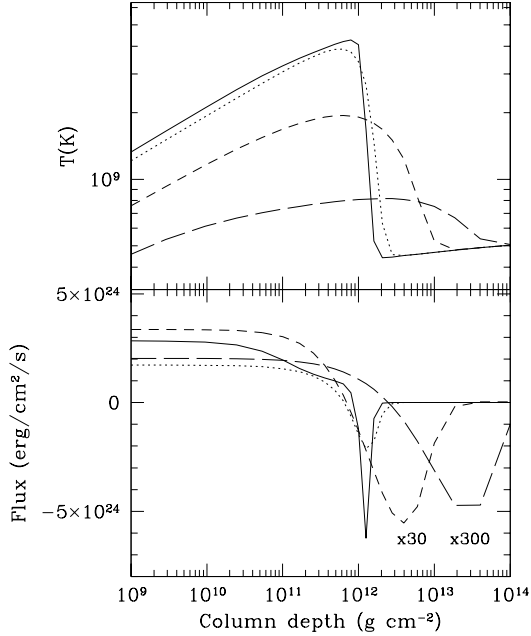}
\caption{Temperature and flux profiles for $y_b=10^{12}\ {\rm g\
cm^{-2}}$, $E_{17}=1$, and $t=$ 10 minutes (solid lines), 1 hour
(dotted lines), 1 day (short-dashed lines), and 10 days (long-dashed
lines) after ignition. In the lower panel, the $t=1$ hour (1 day)
flux profile is shown scaled by a factor of 30 (300).}
\label{fig:go2}
\end{figure}

\begin{figure}
\plotone{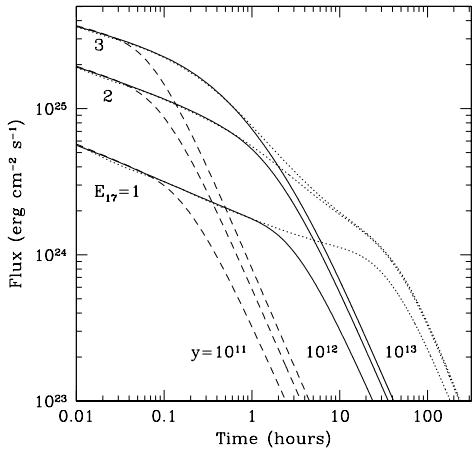}
\caption{Flux from the surface against time for different values for the energy release $E_{17}$ and layer thickness $y_b$ (in ${\rm g\ cm^{-2}}$).}
\label{fig:gop3}
\end{figure}

\begin{figure}
\plotone{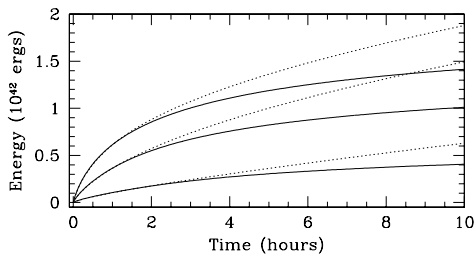}
\caption{Cumulative energy released from the surface as a function of
time, for $y_b=10^{12}$ (solid lines) and $10^{13}\ {\rm g\ cm^{-2}}$
(dotted lines) and in each case (bottom to top) $E_{17}=1$, $2$, and
$3$.}
\label{fig:gopfr}
\end{figure}

To find the temperature profile just after the fuel burns, we deposit an energy $E_\mathrm{nuc}=E_{17} 10^{17}\ {\rm erg\ g^{-1}}$ throughout the layer. Since carbon burning to iron gives $\approx 10^{18}\ {\rm erg\ g^{-1}}$, we expect $E_{17}=1$ to correspond to $X_C\approx 0.05$--$0.1$, depending on how much energy is contributed by photodisintegration (Schatz et al.~2003). At each depth, we calculate the temperature of the layer, $T_f$, from $\int^{T_f}_{T_i} c_P dT = E_{\rm nuc}$, where $T_i$ the initial temperature. Following CB01, a simple analytic estimate of $T_f$ is as
follows. The electrons are degenerate and relativistic for
$\rho\gtrsim 10^7\ {\rm g\ cm^{-3}}$, giving $\rho Y_e = 1.5\times
10^8\ {\rm g\ cm^{-3}}\ P_{26}^{3/4}$, Fermi energy $E_F=2.7\ {\rm
MeV}\ P_{26}^{1/4}$, and pressure scale height $H=y/\rho=67\ {\rm m}\
P^{1/4}_{26}Y_e/g_{14}$, where $P=P_{26}\ 10^{26}\ {\rm erg\ cm^{-3}}$. The heat capacity $c_P$ is determined mainly
by the electrons, $c_P\approx \pi^2(Y_ek_B/m_p)(k_BT/E_F)= 2.6\times 10^7\
{\rm erg\ g^{-1}\ K^{-1}}\ T_9Y_eP^{-1/4}_{26}$. Integrating and
assuming $T_i\ll T_f$, we find the convectively stable temperature profile $T_f = 3.6\times 10^9\ {\rm K}\ E^{1/2}_{17}P^{1/8}_{26}Y_e^{-1/8}$, insensitive to depth and depending mainly on $E_{17}$.

Our thermal evolution code uses the method of lines, in which the
right hand side of equation (1) is differenced over a spatial grid,
and the resulting set of ordinary differential equations integrated
using a stiff integrator. We choose a uniform grid in $\sinh^{-1}(\log
y/y_b)$, which concentrates grid points around the base of the layer
$y_b$, resolving the initial temperature discontinuity. We place the
outer boundary at $y = 10^8\ {\rm g\ cm^{-2}}$, and set flux $\propto
T^4$ there; at the inner boundary, typically $y\approx 10^{14}\ {\rm
g\ cm^{-2}}$, we assume vanishing flux.  We assume the layer is
heated before the flash by a $10^{21}\ {\rm erg\ cm^{-2}\ s^{-1}}$
flux from the crust, and that all the fuel burns to $^{56}$Fe. Our results are not sensitive to the details of the grid, or boundary conditions (for times longer than the thermal
time at the top zone). We calculate the equation of state, opacity, neutrino emissivity, and heat capacity as described by Schatz et al.~(2003).

Figure 1 shows temperature and flux profiles 10 minutes, 1 hour, 1
day, and 10 days after ignition for a model with $E_{17}=1$ and $y_b
= 10^{12}\ {\rm g\ cm^{-2}}$. Figure 2 shows a series of lightcurves
for different $y_b$ and $E_{17}$. At early times, as the
outer parts of the layer thermally adjust, the radiative flux depends
mostly on $E_{17}$. At late times, after the cooling wave reaches the base of the layer, the flux depends mostly on $y_b$, and falls off as a power law $F\propto t^{-4/3}$. Figure 3 shows the cumulative energy release for $y_b = 10^{12}$ and $10^{13}\ {\rm g\ cm^{-2}}$ and $E_{17}=1$--$3$.  In the first few hours, the energy released from the surface is $\approx 10^{42}\ {\rm ergs}$, the exact value being mainly sensitive to $E_{17}$,
rather than depth. A significant fraction of the heat is initially
conducted inwards and released on a longer timescale, as pointed out
by Strohmayer \& Brown (2002). 

The physical reason for the late-time power law flux decay is that as time evolves, the peak of the temperature profile moves to greater depths where the thermal timescale to the surface is longer\footnote{A similar problem is ohmic decay of crustal magnetic
fields, where power law decay is also expected (Sang \& Chanmugam
1987; Urpin, Chanmugam, \& Sang 1994)} (see Fig.~1). The simplest analytic model is
a slab with constant thermal diffusivity $D$, whose temperature is
perturbed close to the surface, for example at a depth $x=a$ (where
$x=0$ is the surface). For a delta-function perturbation initially,
the temperature evolution is given by the Green's function\footnote{A
simple way to obtain this result is to apply the method of images to
the Green's function for an unbounded domain $T(x,t)\propto t^{-1/2}
\exp\left(-x^2/4Dt\right)$. Eichler \& Cheng (1989) derive a similar
result for a power law dependence of conductivity on depth, which also
shows self-similar behavior at late times (see Lyubarsky, Eichler, \&
Thompson 2002 for a recent application to cooling of SGR 1900+14
after an outburst).}
\begin{equation}
T(x,t) = {1\over \sqrt{\pi Dt}} \sinh\left({ax\over 2Dt}\right)
\exp\left(-{x^2+a^2\over 4Dt}\right)
\end{equation}
and the surface flux is $F\propto (\partial T/\partial x)_{x=0}\propto
t^{-3/2}\exp(-\tau/t)$, where $\tau$ is the thermal time at the
initial heating depth $\tau = 4a^2/D$.  For an initial "top hat"
temperature profile $T (x < a) = 1$, $T (x > a) = 0$, the surface flux
is $F\propto (\tau/t)^{1/2}\left[1-\exp(-\tau /t)\right]$. For $t
<\tau$, before the cooling wave reaches the base of the layer,
$F\propto t^{-1/2}$; for $t >\tau$, the solution is independent of
the initial temperature profile, and $F\propto t^{-3/2}$.  

The numerical results show a similar behavior, although with different
power law indices. The relevant timescale in this case is $t_{\rm
cool} = H^2/D$, where $D = K/\rho c_P$.  The electron conductivity is
$K=\pi^2n_ek^2_BT/3m_\star\nu_c$, where $m_\star = E_F/c^2$, and
$\nu_c$ is the electron collision frequency. When electrons dominate
the heat capacity, the thermal diffusivity takes the particularly
simple temperature-independent form $D = c^2/3\nu_c$. For electron-ion
collisions, $\nu_c = 9.3\times 10^{16}\ {\rm s^{-1}}\
P_{26}^{1/4}\langle Z^2/A\rangle\Lambda_{ei}/Y_e$ (e.g.~see Appendix of Schatz et al.~1999), giving
\begin{equation}
t_{\rm cool} = 3.8\ {\rm hrs}\ y^{3/4}_{12} \left({Y_e\langle
Z^2/A\rangle\Lambda_{ei}\over 6}\right) \left({g_{14}\over
2.45}\right)^{-5/4}
\end{equation}
(see also eq.~[10] of CB01), where we insert the appropriate numbers
for $^{56}$Fe composition. The simple ``top hat'' solution for constant conductivity motivates a fit to the numerical solutions,
\begin{equation}\label{eq:fit}
F_{25} = 0.2\ t_{\rm hr}^{-0.2} \ E_{17}^{7/4}
\left[1-\exp\left(-0.63\ t_{\rm cool}^{4/3}E_{17}^{-5/4}t_{\rm
hr}^{-1.13}\right)\right],
\end{equation}
where $t_{\rm hr} = t/1\ {\rm hour}$. For
$t > t_{\rm cool}$, $F_{25}=0.13\ E^{1/2}_{17}(t/t_{\rm
cool})^{-4/3}$. The transition from $F\propto t^{-0.2}$ to $F\propto
t^{-4/3}$ occurs when $t/t_{\rm cool}\approx E_{17}^{-1.1}$. 

Equation (\ref{eq:fit}) fits the numerical results to better than a factor of two for models without substantial neutrino emission. As emphasised by Strohmayer \& Brown (2002), neutrino cooling is important for large carbon fractions: it depresses the flux at $t\approx 5$--$10$ hours for the models with $E_{17}=2$ and $3$, $y =10^{13}\ {\rm g\ cm^{-2}}$ in Figure 2. Whenever neutrinos dominate the cooling, the peak temperature is large enough that emission is by pair annhilation. A good fit to the neutrino energy loss rate is $\epsilon_\nu\approx 10^4\ {\rm erg\ g^{-1}\ s^{-1}} T^{12}_9 y^{-3/2}_{12}$, giving a
cooling time $t_\nu=c_PT/\epsilon_\nu=2.5\times 10^{12}\ {\rm s}\
y^{5/4}_{12}T^{-10}_9$. Inserting the peak temperature from equation
(2) gives $t_\nu\approx 300\ {\rm hrs}\ E^{-5}_{17}$. Neutrinos dominate when $t_\nu<t_{\rm cool}$, or when $E_{17}>2.3\ y^{-3/20}_{12}$.

\section{Comparison to Observations}

The cooling curves in Figure \ref{fig:gop3} compare well with observed
lightcurves, including a rapid initial decay on hour timescales, followed by an extended tail of emission (as observed following some superbursts, e.g.~KS~1731-260, Kuulkers et al.~2002; Ser X-1, Cornelisse et al.~2002). We will present a detailed comparison with the observed superburst lightcurves in a future paper. The initial decay from the peak depends mostly on $E_{17}$, and so it should be possible to constrain the amount of fuel consumed in the superburst. Our models do not resolve the peak itself, since this depends on the details of how the burning propagates out to the surface; however, for $E_{17}\gtrsim 2$--$3$, the flux exceeds the Eddington flux, $F_{\rm Edd}=3\times 10^{25}\ \mathrm{erg\ cm^{-2}\ s^{-1}}/(1+X)$, where $X$ is the H fraction, for timescales of minutes. Superburst peak luminosities are generally less than the Eddington luminosity (Kuulkers 2003), implying $E_{17}\lesssim 2$. The one exception is the superburst from 4U~1820-30, which showed dramatic photospheric radius expansion lasting for several minutes (Strohmayer \& Brown 2002). This is consistent with the proposal that this source, which accretes and burns He rich material, produces large quantities of carbon (Strohmayer \& Brown 2002; Cumming 2003a). The transition to the late-time power law occurs after $t\approx 4\ {\rm h}\ E_{17}^{-1.11}y_{12}^{3/4}$ (eq.~[5]), which corresponds to $F_{25}\approx 0.13 E_{17}^2$. It may therefore be possible to measure the power law decay using superburst tails, although this depends upon being able to subtract out the underlying accretion luminosity, $F_{\rm accr,25}\approx 0.1\ (\dot M/0.1 \dot M_{\rm Edd})$, in a reliable way. 

\begin{figure}
\plotone{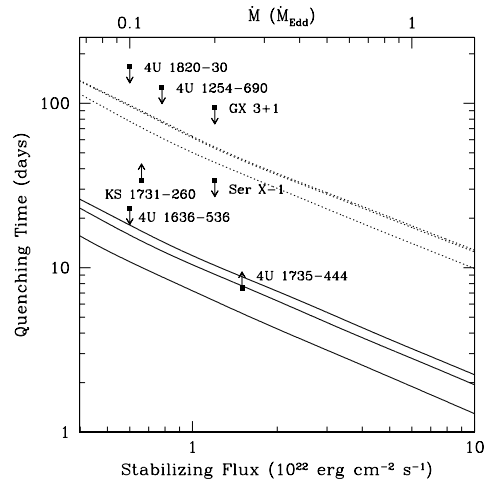}
\caption{Observed and predicted quenching timescales of normal Type I bursts following a superburst. We show the predicted quenching time for $y_b=10^{12}$ (solid lines) and $10^{13}\
{\rm g\ cm^{-2}}$ (dotted lines) as a function of both the critical flux needed to stabilize H/He burning $F_{\rm crit}$, and the accretion rate (given in terms of $F_{\rm crit}$ by eq.~[\ref{eq:fcrit}]). For each value of $y_b$, we show curves for (bottom to top) $E_{17}=1$, $2$ and $3$. The observed upper or lower limits on t$_{\rm quench}$, and the estimated $\dot M$, are taken from Table 1 of Kuulkers 2003, with an updated value for 4U~1636-53 (Kuulkers, private communication). }
\label{fig:quench}
\end{figure}

Another way to probe the late-time cooling is to use the remarkable
observation that Type I bursts disappear (are ``quenched'') for $t_{\rm
quench}\approx$ weeks following the superburst (e.g.~Kuulkers
2003). CB01 proposed that the cooling flux from the superburst
temporarily stabilizes the H/He burning. An estimate of the critical stabilizing flux, $F_{\rm crit}$, is as follows. The condition for temperature fluctuations to grow, and unstable He ignition to occur, is $\nu\epsilon_{3\alpha} = \eta\epsilon_{\rm cool}$ (Fushiki \& Lamb 1987), where $\epsilon_{3\alpha}$ is the triple alpha ($3\alpha$) energy
production rate, $\epsilon_{\rm cool}$ is a local approximation to the
cooling rate, and $\nu$ and $\eta$ are the respective temperature
sensitivities. For a large flux from below, the He burns stably before
reaching this ignition condition, at a depth where the time to
accumulate the layer equals the He burning time, $y/\dot m =
YQ_{3\alpha}/\epsilon_{3\alpha}$, where $Y$ is the He mass fraction,
$\dot m$ is the local accretion rate per unit area, and $Q_{3\alpha} =
5.84\times 10^{17}\ {\rm erg\ g^{-1}} = 0.606$ MeV per nucleon is the
$3\alpha$ energy release. At the transition from
unstable to stable burning, both criteria are satisfied at the base of
the H/He layer. Using the first condition to eliminate
$\epsilon_{3\alpha}$ from the second, and writing $\epsilon_{\rm
cool}\approx F/y$, gives $F=\nu\dot m Q_{3\alpha}Y/\eta=6.2\times 10^{22}\
{\rm erg\ cm^{-2}\ s^{-1}} (\dot m/\dot m_{\rm
Edd})(Y/0.3)(\nu/4\eta)$. Some of this flux is provided by hot CNO
burning of accreted H, $F_H\approx\epsilon_Hy=5.8\times 10^{21}\ {\rm
erg\ cm^{-2}\ s^{-1}} y_8 (Z/0.01)$ (Cumming \& Bildsten 2000; $Z$ is
the metallicity); the remainder is $F_{\rm crit}=F-F_H$. This estimate
agrees well with a more detailed calculation using the ignition models
of Cumming \& Bildsten (2000), in which we find $F_{\rm
crit}\approx\dot m Q_{3\alpha} \approx 0.7\ {\rm MeV}$ per accreted
nucleon, almost independent of $\dot M$. Therefore,
\begin{equation}\label{eq:fcrit}
F_{{\rm crit},22}\approx 6\ (\dot m/ \dot m_{\rm Edd})
\end{equation}
(see also Paczynski 1983a; Bildsten 1995). Equation (5) in the limit $t\gg t_{\rm cool}$ gives
\begin{equation}
t_{\rm quench} = 38\ t_{\rm cool}\ F_{{\rm crit},22}^{-3/4}
E_{17}^{3/8} = 6\ {\rm days}\ y^{3/4}_{12} F^{-3/4}_{{\rm crit},22}
E^{3/8}_{17},
\end{equation}
which gives $t_\mathrm{quench}$ in terms of the thickness of the layer and energy release.

Figure 5 compares the predicted and observed quenching times. The
observations of $t_{\rm quench}$ and accretion rates (used to find
$F_{\rm crit}$ from eq.~[\ref{eq:fcrit}]) are taken from Kuulkers (2003) (except for 4U~1636-53, which has a revised upper limit of $23$ days, Kuulkers private communication). The observations are upper or lower limits only: nonetheless, the general
agreement is very good and supports the quenching picture suggested by
CB01. There is much to learn from a careful comparison of superburst lightcurves and the corresponding quenching times, separately constraining both $E_{17}$ and $y_b$.

\section{Summary and Conclusions}

We have presented the first multi-zone models of the cooling phase of
superbursts. The flux decay is not exponential, but power-law
(eq.~[5]). For $t<t_{\rm cool}$, where $t_{\rm cool}$ is the cooling
time at the base of the layer, the flux depends mostly on the energy release $E_{17}$, and is insensitive to depth: the inwards travelling
cooling wave does not yet know that the layer has a finite thickness. For
$t>t_{\rm cool}$, the flux decays as a power law $F\propto t^{-4/3}$,
independent of the initial temperature profile. The power law decay at
late times gives predicted Type I burst quenching times of weeks
(eq.~[6]), consistent with observational limits. Future
comparisons of both superburst lightcurves and quenching times with
observations will constrain both the thickness of the fuel layer and
the energy deposited, particularly when combined with models of normal
Type I bursts from the same source (Cumming 2003a,b).

There is still much to be done in terms of theory. Perhaps the most important
issues are the physics of the rise (which sets the initial condition
for our simulations), and production of the fuel. Important clues to
the first are the observed precursors to superbursts, which may be
normal Type I bursts ignited by the superburst. Recent progress has
been made on the second, with indications from both theory (Schatz et al.~2003; Woosley et al.~2003) and observations (in\ 't Zand et al.~2003) that stable burning may be required to produce enough
carbon to power superbursts. A self-consistent model of H/He burning,
followed by accumulation and ignition of the ashes may require a
better understanding of the transition from unstable to stable burning
observed in normal Type I bursting (e.g.~Cornelisse et al.~2003).

\acknowledgements 
We thank P.~Arras, E.~Brown, R.~Cornelisse, E.~Kuulkers, G.~Ushomirsky, S.~Woosley, and J.~in't Zand for useful comments and discussions. AC is supported by NASA through Hubble Fellowship grant HF-01138 awarded by the Space Telescope Science Institute, which is operated by the Association of Universities for Research in Astronomy, Inc., for NASA, under contract NAS 5-26555. JM acknowledges support from DOE grant
No.\ DE-FC02-01ER41176 to the Supernova Science Center/UCSC.

\end{document}